\documentclass[conference, 10pt]{IEEEtran}

\usepackage{graphicx}
\usepackage{color}
\usepackage{placeins}
\usepackage{float}
\usepackage{tabularx,colortbl}
\usepackage{psfrag}
\usepackage{subfigure}
\usepackage{amsmath}
\usepackage{setspace}

\usepackage{mathtools}

\usepackage{cuted}
\begin{document}
%
\title{Amplitude-Equalized Microwave C-Section}

\author{\IEEEauthorblockN{Shulabh Gupta}
\IEEEauthorblockA{
\begin{minipage}{10.5cm}
\centering
Department of Electronics, Carleton University, 1125 Colonel by Drive, Ottawa, Ontario, Canada, shulabh.gupta@carleton.ca
\end{minipage}}}

\maketitle

\begin{abstract}
An active microwave C-section is proposed which provides a flat magnitude transmission in a wide frequency band along with a frequency-dependent group delay response considering practical dissipation losses. The key lies in integrating a constant gain amplifier inside a microwave C-section, which perfectly compensates the distributed conductor and dielectric losses of the coupler, while preserving the intrinsic dispersion of the C-section. The operation of the proposed device is confirmed using numerical analysis.
\end{abstract}

\IEEEpeerreviewmaketitle

\section{Introduction}

A C-section represents a fundamental building unit of microwave dispersive devices such a wide-band phase shifters \cite{Cristal_TMTT_01_1969}\cite{Schiffman_MTT_1955}\cite{DiffPhaseShifterCPLLines}, equalizers \cite{MYJ_AHP_1980} and real-time signal processors \cite{Gupta_IJCTA_2013}\cite{Generalized_Phasers}. Recently, they have been used extensively to build dispersive devices exhibiting complex group delay profiles, owing to their all-pass transmission properties, which have found important applications in Real-Time Analog Signal Processing (R-ASP) technology \cite{Caloz_MM_2012}. 

A microwave C-section is  2-port device and is based on contra-directional TEM coupled-line couplers \cite{MYJ_AHP_1980}. Under \emph{ideal lossless conditions,} it provides a flat transmission [$|S_{21}(\omega)|=1$] with a strongly dispersive frequency-dependent group delay response [$\tau = \tau(\omega)$]. However, all practical devices exhibit dissipation losses, and therefore, a flat transmission is never achieved in practice, i.e. $|S_{21}(\omega)| \ne1$. Its magnitude transmission is always dependent on the dispersion response following Kramers-Kronig relations and the corresponding Bode gain-phase relations \cite{Gupta_PDM_Arxiv}.  This frequency-dependent magnitude transmission subsequently leads to undesired signal distortion at the output of the corresponding dispersive devices.

To combat this issue, the concept of a \emph{perfect dispersive medium} (PDM) was recently developed in \cite{Gupta_PDM_Arxiv} to decouple the magnitude and phase response of an electromagnetic device, while satisfying all causality requirements, i.e. achieving $|S_{21}(\omega)| = 1$ and $\tau = \tau(\omega)$, simultaneously. The 1-D counterpart of a PDM was demonstrated using microwave C-sections in \cite{Zou_Active_Arxiv}, where each super-cell consisted of two separate loss and gain C-sections leading to amplitude-equalization of the super-cell and dispersion reconfigurability. While dispersion reconfigurability is specific to certain systems such as dispersion code multiple access (DCMA) \cite{DCMA_Gupta_APS}, amplitude equalization is a desired characteristic of any other dispersion based system such as real-time Fourier transformers \cite{Jannson_OL_04_1983}\cite{Laso_TMTT_03_2003}, spectrum sniffers \cite{Nikfal_MWCL_11_2012} or frequency discriminators \cite{Nikfal_TMTT_06_2011}.

In this work, the same flat transmission response and frequency-dependent group delay are achieved simultaneously using a \emph{single} microwave C-section. It will be shown that the distributed dielectric and conductor losses of the C-section can be perfectly compensated in a wide band by integrating a gain-load inside a microwave C-section.  

\section{Conventional Lossy C-section}

A microwave C-section is a 2-port network formed by interconnecting the through and the isolated ports of a 4-port contra-directional coupled-line coupler exhibiting infinite isolation and perfect matching. Its 2-port transfer function can be derived in terms of the known coupled $a(\theta)$ and through $b(\theta)$ transfer functions of the coupler, following the wave-interference approach as \cite{Gupta_TMTT_12_2012}\cite{Gupta_TMTT_09_2010}

\begin{equation}
S_{21}(\theta) = a(\theta) + \frac{ b^2(\theta)}{1- a(\theta)}.\label{Eq:CSecWI}
\end{equation}

\begin{figure}[htbp]
\begin{center}
\psfrag{a}[r][c][0.8]{$v_\text{in}(\omega)$, Port 1}
\psfrag{b}[r][c][0.8]{$v_\text{c}(\omega)$, Port 2}
\psfrag{c}[c][c][0.8]{$\ell = \lambda/4$@$\omega_0$}
\psfrag{d}[l][c][0.8]{$a(\theta=\gamma\ell)$}
\psfrag{e}[c][c][0.8]{$b(\theta =\gamma\ell)$}
\psfrag{f}[l][c][0.8]{$v_\text{th}(\omega)$, Port 3}
\psfrag{g}[l][c][0.8]{$v_\text{iso}(\omega)$, Port 4}
\includegraphics[width=0.6\columnwidth]{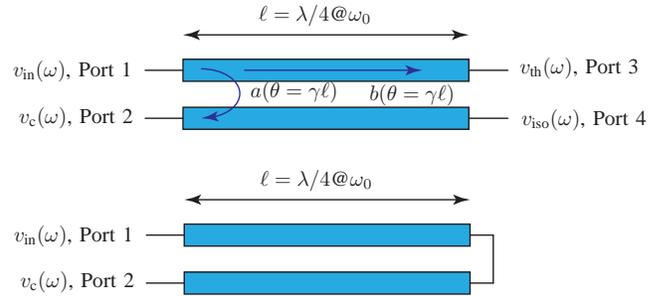}
\caption{A typical configuration of a 2-port microwave C-section formed using a 4-port contra-directional coupled-line coupler, by interconnecting the through and isolated ports of the coupler. The coupler is assumed to have perfect matching and infinite isolation.}\label{Fig:CSection}
\end{center}
\end{figure}

\noindent This configuration is shown in Fig~\ref{Fig:CSection}. The closed form expressions of the coupling and through transfer functions $a$ and $b$, are \cite{Mongia_book_Couplers}

\begin{subequations}
\begin{equation}
a(\theta) = \frac{jk\sin\theta}{\sqrt{1-k^2} \cos\theta + j\sin\theta},
\end{equation}
\begin{equation}
b(\theta) = \frac{\sqrt{1-k^2}}{\sqrt{1-k^2} \cos\theta + j\sin\theta},
\end{equation}\label{Eq:CplThr}
\end{subequations}

\noindent where $k$ is the voltage coupling coefficient and Re$\{\theta(\omega)\} = \gamma(\omega)\ell$ is the electrical length of the transmission with $\gamma$ as the complex propagation constant of the isolated line. The complex propagation constant $\gamma(\omega) = \beta(\omega) - j\alpha(\omega)$, where $\beta$ and $\alpha$ are the per-unit propagation and attenuation constants of the isolated transmission lines, respectively. Substituting the above expressions in \eqref{Eq:CSecWI}, the resulting transfer function is given by

\begin{equation}
S_{21}(\theta) = \left(\frac{1 - j\rho\tan\theta}{1 + j\rho\tan\theta}\right)
\end{equation}

\noindent where $\rho = \sqrt{(1-k)/(1+k)}$.

If the C-section is lossless ($\alpha = 0$), it can be verified that $|S_{21}(\omega)| = 1$, and this is classically referred to as an \emph{all-pass response}, where all the frequencies are transmitted equally in magnitude \cite{MYJ_AHP_1980}. However, this is just a mathematical idealization, as any practical C-section exhibits both conductor and dielectric losses corresponding to a non-zero $\alpha$. For a practical C-section with $\alpha\ne 0$, the transfer function can be re-written as \cite{Zou_Active_Arxiv}

\begin{equation}
S_{21}(\omega) = \left\{\frac{1- \rho\tanh\alpha\ell -j(\rho - \tanh\alpha\ell)\tan\beta\ell}{1+ \rho\tanh\alpha\ell +j(\rho + \tanh\alpha\ell)\tan\beta\ell}\right\},\label{Eq:LossyCsection}
\end{equation}

\noindent with

\begin{equation}
|S_{21}(\omega)| = \sqrt{\frac{(1- \rho\tanh\alpha\ell)^2 + (\rho - \tanh\alpha\ell)^2\tan^2\beta\ell}{(1+ \rho\tanh\alpha\ell)^2 -(\rho + \tanh\alpha\ell)^2\tan^2\beta\ell}}.
\end{equation}

\noindent In this case, two main observations can be made:

\begin{enumerate}
\item The transmission magnitude $|S{21}(\omega)| \ne 1$, and is in fact a strong function of frequency.
\item Minimum transmission, min$\{|S{21}(\omega)|\}$, is at the resonant frequency $\omega_0$ of the coupler where its length $\ell$ is a quarter wavelength long, i.e.

\begin{equation}
\text{min}\{|S{21}(\omega)|\} = \sqrt{\frac{(1- \rho\tanh\alpha\ell)^2 + (\rho - \tanh\alpha\ell)^2}{(1+ \rho\tanh\alpha\ell)^2 -(\rho + \tanh\alpha\ell)^2}}.\label{Eq:S21min}
\end{equation}

\end{enumerate}

A C-section also exhibits a strong dispersion (frequency dependent group delay). Its group delay response $\tau(\omega) = -d\angle S_{21}(\omega)/d\omega$. It can be easily verified that the maximum delay $\tau$ occurs at the resonant frequency $\omega_0$ of the coupler, and is given by

\begin{equation}
\text{max}\{\tau(\omega)\} = \frac{2}{\rho}\frac{d\beta(\omega)}{d\omega}\ell.\label{Eq:TauMax}
\end{equation}

\begin{figure}[htbp]
\begin{center}
\psfrag{a}[c][c][0.8]{transmission $|S_{21}|$ (dB)}
\psfrag{b}[c][c][0.8]{frequency $\omega$ (rad/s)}
\psfrag{c}[c][c][0.8]{delay ($\tau\times2\omega_0/\pi$)}
\psfrag{d}[c][c][0.7]{$1^\text{st}$}
\psfrag{e}[c][c][0.7]{$2^\text{nd}$}
\psfrag{f}[c][c][0.7]{$\Delta\tau$}
\psfrag{g}[c][c][0.7]{$\Delta S $}
\psfrag{A}[c][c][0.8]{$\omega_0$}
\psfrag{B}[c][c][0.8]{$2\omega_0$}
\psfrag{C}[c][c][0.8]{$3\omega_0$}
\psfrag{D}[c][c][0.8]{$4\omega_0$}
\psfrag{x}[c][c][0.8]{$1$}
\psfrag{y}[c][c][0.8]{$2$}
\includegraphics[width=0.65\columnwidth]{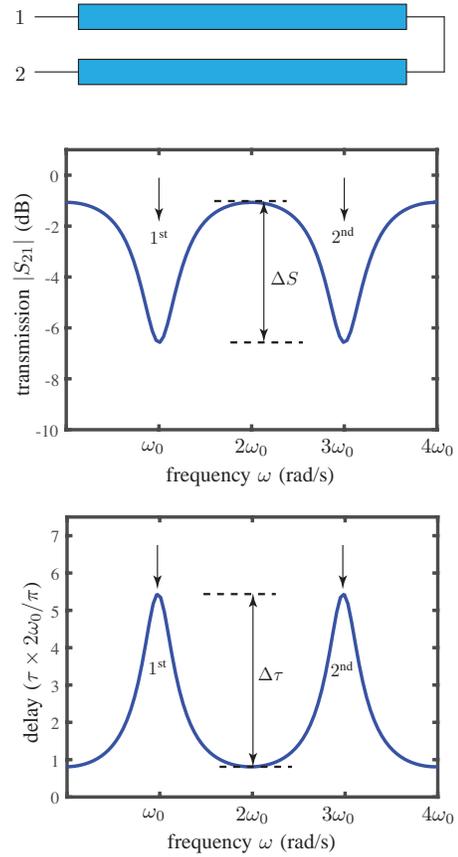}
\caption{The magnitude transmission and delay responses of a passive microwave C-section of Fig.~\ref{Fig:CSection} computed using \eqref{Eq:LossyCsection}. The design parameters are: $k=0.71$, $\alpha\ell = 0.15$}\label{Fig:CSectionTD}
\end{center}
\end{figure}

\noindent Therefore, to achieve a larger delay swing $\Delta\tau$, $\rho$ should be reduced, i.e. larger coupling $k$. However, following \eqref{Eq:S21min}, a smaller $\rho$ also leads to lower signal transmission, i.e. large $\Delta S$. Therefore, stronger the dispersion, more amplitude distortion the device will exhibit. This is consistent from the physical perspective, where the frequency component with the largest group delay propagates longer inside the structure, and is attenuated the most. Fig.~\ref{Fig:CSectionTD} shows the typical amplitude and delay response of a practical microwave C-section. The lowest transmission and the maximum group delay, both occur at the fundamental resonant frequency and all its harmonics.

\section{Proposed Amplitude-Equalized Active C-section}

As mentioned in the introduction, a frequency dependent transmission magnitude is a problematic of a C-section. The only possibility of a flat magnitude transmission $|S_{21}(\omega)|=1$ in a practical passive C-section is a trivial solution of $k=0$. This corresponds to a \emph{constant group delay}, rendering the device useless for any signal processing application. So how do we achieve a flat magnitude transmission from a microwave C-section preserving its strongly dispersive characteristic?  

\begin{figure}[htbp]
\begin{center}
\psfrag{a}[r][c][0.8]{$v_\text{in}(\omega)$, Port 1}
\psfrag{b}[r][c][0.8]{$v_\text{out}(\omega)$, Port 2}
\psfrag{c}[c][c][0.8]{$\ell = \lambda/4$@$\omega_0$}
\psfrag{d}[l][c][0.8]{$a(\theta=\gamma\ell)$}
\psfrag{e}[c][c][0.8]{$b(\theta =\gamma\ell)$}
\psfrag{f}[c][c][0.8]{$G$}
\includegraphics[width=0.6\columnwidth]{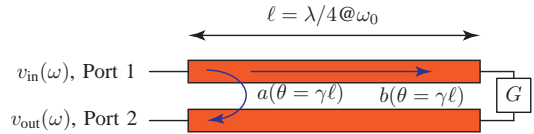}
\caption{Proposed configuration of an amplitude-equalized microwave C-section, where the through and isolated ports of the corresponding coupler is connected through a load $G$.}\label{Fig:ActiveCsection}
\end{center}
\end{figure}

To answer this question, let us modify the microwave C-section where the through and the isolated ports of the corresponding coupler are now connected through a matched load $G$, as shown in Fig.~\ref{Fig:ActiveCsection}. Using the wave-interference approach, the transfer function of such a C-section is given by

\begin{equation}
S_{21}(\theta) = a(\theta) + \frac{G\cdot b^2(\theta)}{1- a(\theta)\cdot G},\label{Eq:ActCSecWI}
\end{equation}

\noindent which when $G=1$, corresponds to a passive C-section, and reduces to \eqref{Eq:LossyCsection}, with $|S_{21}(\omega)|\ne1$.

\begin{figure}[htbp]
\begin{center}
\psfrag{a}[c][c][0.8]{transmission $|S_{21}|$ (dB)}
\psfrag{b}[c][c][0.8]{frequency $\omega$ (rad/s)}
\psfrag{c}[c][c][0.8]{delay ($\tau\times2\omega_0/\pi$)}
\psfrag{d}[c][c][0.8]{$\Delta S = 0$}
\psfrag{e}[c][c][0.8]{$\boxed{G = 1.13}$ \eqref{Eq:OptimumGain}}
\psfrag{A}[c][c][0.8]{$\omega_0$}
\psfrag{B}[c][c][0.8]{$2\omega_0$}
\psfrag{C}[c][c][0.8]{$3\omega_0$}
\psfrag{D}[c][c][0.8]{$4\omega_0$}
\psfrag{x}[c][c][0.7]{$1$}
\psfrag{y}[c][c][0.7]{$2$}
\psfrag{f}[c][c][0.7]{$G$}
\includegraphics[width=0.65\columnwidth]{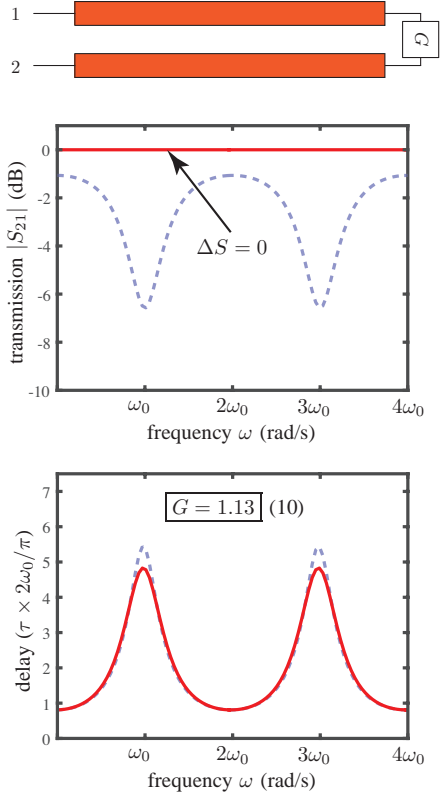}
\caption{The magnitude transmission and delay responses of an active microwave C-section of Fig.~\ref{Fig:ActiveCsection} computed using \eqref{Eq:ActCSecWI}. The responses of a passive C-section is also shown using dashed curves for comparison. The design parameters are: $k=0.71$, $\alpha\ell = 0.15$ and the gain $G_0$ calculated using \eqref{Eq:Condition}.}\label{Fig:ActiveCSectionTD}
\end{center}
\end{figure}

This expression suggests that the load parameter $G$ acts as a free parameter which maybe used to compensate for the frequency-dependent losses in a practical C-section. Let us intuitively enforce the amplitude-equalization at the fundamental resonant frequency $\omega_0$ ($\beta\ell=\pi/2$) of the coupler, i.e.

\begin{equation}
a(\pi/2 - j\alpha\ell) + \frac{G\cdot b^2(\pi/2 - j\alpha\ell)}{1- a(\pi/2 - j\alpha\ell)\cdot G} = \pm 1, \label{Eq:Condition}
\end{equation}

\noindent so that $|S_{21}(\omega_0)|=1$. The $\pm$ signs on the RHS corresponds to resonant or anti-resonant conditions exhibiting the two group delay extrema.  Using \eqref{Eq:CplThr} in \eqref{Eq:Condition}, this equation can be solved for an optimum value of $G = G_0$, resulting in the following value:

\begin{equation}
\frac{(k+2)\{(1- k)\cosh^2(\alpha\ell)+\sqrt{1- k^2}\sinh(2\alpha\ell)/2\}+(k^2-1)}{k(1 + k)\cosh^2(\alpha\ell) + k\sqrt{1- k^2}\sinh(2\alpha\ell)/2 - (k^2-1)}.
 \label{Eq:OptimumGain}
\end{equation}

Figure~\ref{Fig:ActiveCSectionTD} shows the modified magnitude and delay responses of the amplitude-equalized C-section, following the optimum value of $G_0$. The transmission magnitude is now perfectly equalized to $0$~dB. And remarkably, the C-section maintains a strong dependence of the group delay on frequency, as desired.

Following interesting observations and conclusions can be made from this result:

\begin{enumerate}
\item The optimum value $G_0 > 1$, and thus represents a gain load. This is expected as an active element is naturally required to compensate for the dissipation in the C-section. Thus a practical C-section with a flat transmission and finite dispersion is an \emph{active C-section}.
\item A \emph{frequency-independent value} of $G_0$ is found to be sufficient to equalize the frequency-dependent magnitude transmission of a C-section in a wide-bandwidth. This is, a priori, favourable for practical realizations, where the amplifier is needed to provide a constant gain within the frequency band of interest.
\item The optimum gain $G_0 \propto 1/k$, i.e. larger gain is required for low-coupling coefficients $k$. Recalling from \eqref{Eq:TauMax} that the delay $\tau$ is proportional to $k$, a smaller gain $G_0$ is required in more dispersive devices. A typical plot of $G_0$ vs coupling coefficient $k$ is shown in Fig.~\ref{Fig:G0vsK}. This feature is particularly beneficial in strongly dispersive devices, typically required in R-ASP applications, thereby putting moderate requirements on the realizable amplifier gain.

\begin{figure}[htbp]
\begin{center}
\psfrag{a}[c][c][0.8]{optimum Gain $G_0$}
\psfrag{b}[c][c][0.8]{coupling coefficient $k$}
\psfrag{x}[c][c][0.7]{$1$}
\psfrag{y}[c][c][0.7]{$2$}
\psfrag{f}[c][c][0.6]{$G$}
\psfrag{c}[c][c][0.7]{$G = 1.13$}
\psfrag{d}[c][c][0.7]{$k = 0.71$}
\includegraphics[width=0.75\columnwidth]{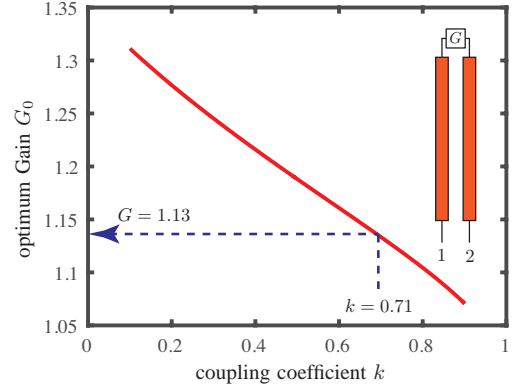}
\caption{Relation between the optimum gain value $G_0$ and the coupling coefficient $k$ in the microwave C-section using \eqref{Eq:OptimumGain} for fixed $\alpha\ell = 0.15$.}\label{Fig:G0vsK}
\end{center}
\end{figure}

\item The optimum gain value is directly proportional to the attenuation factor $\alpha\ell$, as shown in Fig.~\ref{Fig:G0vsAlpha}. The gain increases monotonically with loss, consistent with the physical perspective that more gain needs to be injected into the C-section to compensate for the total dissipation losses.   

\begin{figure}[htbp]
\begin{center}
\psfrag{a}[c][c][0.8]{optimum Gain $G_0$}
\psfrag{b}[c][c][0.8]{per-unit attenuation $\alpha\times\ell$}
\psfrag{x}[c][c][0.7]{$1$}
\psfrag{y}[c][c][0.7]{$2$}
\psfrag{f}[c][c][0.6]{$G$}
\psfrag{c}[c][c][0.7]{$G = 1.13$}
\psfrag{d}[c][c][0.7]{$\alpha\ell = 0.15$}
\includegraphics[width=0.75\columnwidth]{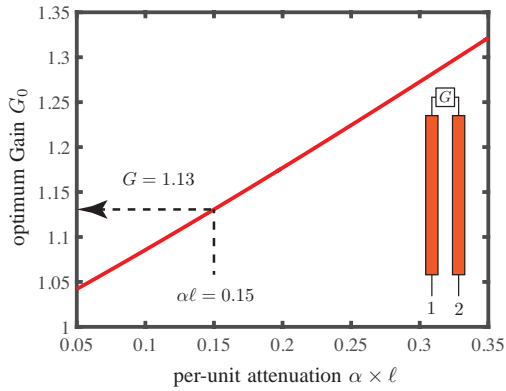}
\caption{Relation between the optimum gain value $G_0$ and the attenuation factor $\alpha\ell$ in the microwave C-section using \eqref{Eq:OptimumGain} for a fixed coupling coefficient $k = 0.71$.}\label{Fig:G0vsAlpha}
\end{center}
\end{figure}

\item A reduction in the group delay swing is observed compared to its passive counterpart, and this appears to be a potential drawback of the proposed approach.

\end{enumerate}

This example thus confirms that an actively loaded C-section is capable of providing a flat magnitude transmission with dispersion leading to a true all-pass response in practice, taking all dissipation into account. This single active C-section thus acts as a fundamental unit which can then be used to synthesize more complex delay responses by cascading several of them with different resonant frequencies and coupling coefficients \cite{Gupta_IJCTA_2013}. By properly tuning the gain responses of each of the C-section, the overall magnitude transmission can be equalized following the proposed technique for a specified dispersion response.  

\section{Conclusions}

An active microwave C-section has been proposed which provides a flat magnitude transmission in a wide frequency band along with a frequency-dependent group delay response. The key lies in integrating a constant gain amplifier inside a microwave C-section, which perfectly compensates the distributed conductor and dielectric losses of the coupler, while preserving the intrinsic dispersion of the C-section. The proposed amplitude-equalized active C-section is compatible with standard MMIC processes \cite{MMIC_Book} and may find practical applications in wide range of phase shifting and dispersive devices for microwaves. Moreover, the proposed concept is also compatible with micro-ring resonators \cite{MicroringReview} which are based on co-directional coupled-line couplers, suitable for high-frequency devices and compatible with silicon photonic technologies for optics and THz applications \cite{SiliconPhotonics_Review_WY}. The detailed stability analysis and experimental demonstration of the proposed active C-section is currently under progress and will be reported elsewhere.

\bibliographystyle{IEEETran}
\bibliography{2016_Gupta_Active_CSection_APRASC}

\end{document}